\documentstyle[twoside,fleqn,espcrc2]{article}
   \title{Quasi-nuclear and quark model baryonium: historical survey
  \thanks{Dedicated to the memory of C.B. Dover and I.S. Shapiro}
\thanks{Invited talk at the Euroconference QCD99, to appear in the 
Proceedings, ed.\ S.~Narison}
\thanks{Preprint ISN 99-95, nucl-th/9909030}
  }
    \author{Jean-Marc Richard 
    \address{
    Institut des Sciences Nucl\'eaires, Universit\'e Joseph 
    Fourier--CNRS-IN2P3\\
    53, avenue des Martyrs, F-38026 Grenoble Cedex, France}}
\def\N{\hbox{N}}
\def\NN{{\N \N}}
\def\Nbar{{\overline{\N}}}
\def\NNbar{{\N\Nbar}}
\def\p{\hbox{p}}
\def\pbar{{\bar{\p}}}
\def\n{\hbox{n}}
\def\nbar{{\bar{\n}}}
\def\q{\hbox{q}}
\def\qbar{{\bar{\q}}}
\def\qqbar{{\q\qbar}}
\def\Q{\hbox{Q}}

\def\SLJ#1#2#3{ {}^{#1}{\hbox{#2}}_{#3}}
\begin{document}
  
\begin{abstract}\noindent
    We review ideas and speculations concerning possible bound 
    states or resonances coupled to the nucleon--antinucleon channel.
\end{abstract}
 \maketitle
\section{Introduction}
\label{Introduction}
In the 70's, there has been many indications of new mesons coupled to 
the nucleon--antinucleon ($\NNbar$) system. States below the $\NNbar$ 
threshold were claimed, e.g., in radiative transitions $\NNbar\to 
X+\gamma$. Above the threshold, states were tentatively seen as bumps in cross 
sections.

Clarifying the experimental situation was one of the main motivation 
for building the low-energy part of the antiproton beam facility at 
CERN. Most candidates for baryonium have not been confirmed by careful scans. It 
remains, however,
\begin{itemize}
    \item Some of the multimeson states seen below threshold in 
    annihilation experiments might have to do with baryonium.  This is 
    the case in particular for the state called ``AX'' (now more 
    prosaically $f_{2}(1565)$  \cite{Caso:1998tx}), as pointed out by 
    Dover  \cite{Dover:1990kn}.
    \item Evidence for broad baryonium states was based on elastic 
    scattering and annihilation into two mesons.  See, e.g., 
    \cite{Carter:1978ha} and references therein.  The PS172 
    collaboration at LEAR has measured the differential cross section 
    and analysing power of $\NNbar$ annihilation into two pseudoscalar 
    mesons at various energies.  Analysis, again, confirms a rich 
    resonance structure \cite{Martin:1980rf}.
    \item There is an intriguing activity in total cross sections, 
    especially for isospin $I=1$ $\NNbar$  \cite{Bertin:1996zh} and for 
    the strangeness-exchange reaction 
    $\NNbar\to\overline{\Lambda}\Lambda$  \cite{Barnes:1994bh}.  A 
    closer look reveals a P-wave enhancement which might be of 
    resonant nature.  Unfortunately, the analysis is not yet published 
    and some of the early claims have not been confirmed in more recent 
    runs.
    \item In the scalar channel ($\SLJ 3P0$ according to the 
    conventional spectroscopic notation), the shift of protonium is 
    slightly larger than expected  \cite{Gotta:1999}. This unavoidably reminds us that a 
    bound state close to threshold in the nuclear spectrum strongly 
    distorts the pattern of atomic levels
\end{itemize}

In short, the intense activity in the hadron spectrum around $2\;$GeV 
makes it difficult to conclude that baryonium is completely dead.  Of 
course, the fashion has evolved: a state that would have been easily 
described years ago as a ``baryonium candidate'' would now 
preferentially be compared to predictions for glueballs or hybrids.  
It remains useful to recall and update the theoretical speculations inspired by 
the baryonium candidates in the late 70's and early 80's.

\section{Quasi-nuclear baryonium}
\label{se:QNB}
\subsection{Brief history}
\label{subse:hist}
The question of possible nucleon--antinucleon ($\NNbar$) bound states 
was raised many years ago, in particular by Fermi and 
Yang \cite{Fermi:1959sa}, who remarked on the strong attraction at large 
and intermediate distances between $\N$ and $\Nbar$.

In the sixties, explicit attempts were made to describe the spectrum 
of ordinary mesons ($\pi$, $\rho$, etc.)  as $\NNbar$ states, an 
approximate realisation of the ``bootstrap'' ideas.  It was 
noticed \cite{Ball:1966sa}, however, that the $\NNbar$ picture hardly 
reproduces the observed patterns of the meson spectrum, in particular 
the property of ``exchange degeneracy'': for most quantum numbers, the 
meson with isospin $I=0$ is nearly degenerate with its $I=1$ partner, 
one striking example being provided by $\omega$ and $\rho$ vector 
mesons.

	In the 70's, a new approach was pioneered by 
	Shapiro \cite{Shapiro:1978wi}, Dover \cite{Dover:1975} and 
	others: in their view, $\NNbar$ states were no  more associated with 
	``ordinary'' light mesons, but instead with new types of mesons 
	with  a mass near the $\NNbar$ threshold and 
	specific decay properties.
	
	This new approach was encouraged by evidence from many intriguing 
	experimental investigations in the 70's, which also stimulated a 
	very interesting activity in the quark model: exotic 
	multiquark configurations were studied, followed by glueballs 
	and hybrid states, more generally all ``non-$\qqbar$'' mesons 
	which will be extensively discussed at this conference.
	
	Closer to the idea of quasi-nuclear baryonium are the 
	meson--meson molecules.  Those were studied mostly by particle 
	physicists, while $\NNbar$ states remained more the domain of 
	interest of nuclear physicists, due to the link with nuclear 
	forces.
\subsection{The $\mathbf{G}$-parity rule}
\label{subse:G-rule}
In QED, it is well-known that the amplitude of $\mu^{+} e^+$ 
scattering, for instance, is deduced from the $\mu^{+} e^-$ one by the 
rule of $C$ conjugation: the contribution from one-photon exchange 
($C=-1$) flips the sign, that of two photons ($C=+1$) is unchanged, etc. 
In short, if the amplitude is split into two parts according to 
the $C$ content of the $t$-channel reaction $\mu^{+}\mu^-\to 
e^{+}e^-$, then
\begin{eqnarray}
    &&{\cal M}(\mu^{+} e^+)={\cal M}_{+}+{\cal M}_{-},\nonumber\\
    &&{\cal M}(\mu^{+} e^-)={\cal M}_{+}-{\cal M}_{-}.
 \end{eqnarray}
The same rule can be formulated for strong interactions and applied to 
relate $\pbar\p$ to $\p\p$, as well as $\nbar\p$ to $\n\p$.  However, 
as strong interactions are invariant under isospin rotations, it is 
more convenient to work with isospin eigenstates, and the rule becomes 
the following.  If the $\NN$ amplitude of $s$-channel isospin $I$ is 
split into $t$-channel exchanges of $G$-parity $G=+1$ and exchanges with 
$G=-1$, the former contributes exactly the same to the $\NNbar$ 
amplitude of same isospin $I$, while the latter changes sign.

This rule is often expressed in terms of one-pion exchange or 
$\omega$-exchange having an opposite sign in $\NNbar$ with respect 
to $\NN$, while $\rho$ or $\epsilon$ exchange contribute with the 
same sign. It should be underlined, however, that the rule is valid much 
beyond the one-boson-exchange approximation. For instance, a crossed 
diagram with two pions being exchanged contributes with the same sign 
to $\NN$ and $\NNbar$.
\subsection{Properties of the $\mathbf{\NN}$ potential}
\label{NNpot}
Already in the early 70's, a fairly decent understanding of long- and 
medium-range nuclear forces was achieved. First, the tail is dominated 
by the celebrated Yukawa term, one-pion exchange, which is necessary 
to reproduce the peripheral phase-shifts at low energy as well as the 
quadrupole deformation of the deuteron \cite{Lacombe:1980dr}.

At intermediate distances, pion exchange, even when supplemented by 
its own iteration, does not provide enough attraction.  It is 
necessary to introduce a spin-isospin blind attraction, otherwise, one 
hardly achieves binding of the known nuclei.  This was called 
$\sigma$-exchange or $\epsilon$-exchange, sometimes split into two 
fictitious narrow mesons to mimic  the large width of this meson, 
which results in a variety of ranges.  The true nature of this meson 
has been extensively discussed in the session chaired by Lucien 
Montanet at this Workshop.  Refined models of nuclear forces describe 
this attraction as due to two-pion exchanges, including the 
possibility of strong $\pi\pi$ correlation, as well as excitation 
nucleon resonances in the intermediate states.  The main conceptual 
difficulty is to avoid double counting when superposing $s$-channel 
type of resonances and $t$-channel type of exchanges, a problem known 
as ``duality''.

To describe the medium-range nuclear forces accurately, one also needs 
some spin-dependent contributions.  For instance, the P-wave 
phase-shifts with quantum numbers $\SLJ{2S+1}{L}{J}=\SLJ3P0 $, 
$\SLJ3P1$ and $\SLJ3P2$, dominated at very low energy by pion 
exchange, exhibit different patterns as energy increases.  Their 
behaviour is typical of the spin-orbit forces mediated by vector 
mesons.  This is why $\rho$-exchange and to a lesser extent, 
$\omega$-exchange cannot be avoided.

Another role of $\omega$-exchange is to moderate somewhat the 
attraction due to two-pion exchange.  By no means, however,  can it 
account for the whole repulsion which is observed at short-distances, 
and which is responsible of the saturation properties in 
heavy nuclei and nuclear matter.  In the 70's, the short-range $\NN$ 
repulsion was treated empirically by cutting off or regularising the 
Yukawa-type terms due to meson-exchanges and adding some {\sl ad-hoc} 
parametrization of the core, adjusted to reproduce the S-wave 
phase-shifts and the binding energy of the deuteron.

Needless to say, dramatic progress in the description of nuclear forces 
have been achieved in recent years.  On the theory side, we 
understand, at least qualitatively, that the short-range repulsion is 
due to the quark content of each nucleon.  This is similar to the 
repulsion between two Helium atoms: due to the Pauli principle, the 
electrons of the first atom tend to expel the electrons of the second 
atom.  On the phenomenological side, accurate models such as the 
Argonne potential \cite{Wiringa:1995wb} are now used for sophisticated 
nuclear-structure calculations.
\subsection{Properties of the $\mathbf{\NNbar}$ potential}
\label{NNbarpot}
What occurs if one takes one of the $\NN$ potentials available in the 
70's, such as the Paris potential \cite{Lacombe:1980dr} or one the many 
variants of the one-boson-exchange models \cite{Erkelenz:1974uj}, and 
applies to it a $G$-parity transformation?  The resulting $\NNbar$ 
potential exhibits the following properties:

{\em 1)} $\epsilon$ (or equivalent) and $\omega$ exchanges, which 
partially cancel  each other in the $\NN$ case, now add up coherently. 
This means that the $\NNbar$ potential is, on the average, deeper 
than the $\NN$ one. As the latter is attractive enough to bind the 
deuteron, a rich spectrum can be anticipated for $\NNbar$.

{\em 2)} The channel dependence of $\NN$ forces is dominated by a 
strong spin-orbit potential, especially for $I=1$, i.e., 
proton--proton.  This is seen in the P-wave phase-shifts, as mentioned 
above, and also in nucleon--nucleus scattering or in detailed 
spectroscopic studies.  The origin lies in coherent contributions from 
vector exchanges ($\rho$, $\omega$) and scalar exchanges (mainly 
$\epsilon$) to the $I=1$ spin-orbit potential.  Once the $G$-parity 
rule has changed some of the signs, the spin-orbit potential becomes 
moderate, in both $I=0$ and $I=1$ cases, but one observes a very 
strong $I=0$ tensor potential, due to coherent contributions of 
pseudoscalar and vector exchanges \cite{Dover:1978br}.  This property 
is independent of any particular tuning of the coupling constants and 
thus is shared by all models based on meson exchanges.

\subsection{Uncertainties on the $\mathbf{\NNbar}$ potential}   
\label{Uncert}
Before discussing the bound states and resonances in the $\NNbar$ 
potential, it is worth recalling some limits of the approach.

{\em 1)} There are cancellations in the $\NN$ potential.  If a 
component of the potential is sensitive to a combination 
$g_{1}^{2}-g_{2}^{2}$ of the couplings, then a model with $g_{1}$ and 
$g_{2}$ both large can be roughly equivalent to another where they are 
both small.  But these models can substantially differ for the $\NNbar$ 
analogue, if it probes the combination $g_{1}^{2}+g_{2}^{2}$.

{\em 2)} In the same spirit, the $G$-parity content of the $t$-channel 
is not completely guaranteed, except for the pion tail.  In 
particular, the effective $\omega$ exchange presumably incorporates 
many contributions besides some resonating three-pion exchange.

{\em 3)} The concept of $\NN$ potential implicitly assumes the 
6-quark wave function is factorised into two nucleon-clusters $\Psi$ 
and a relative wave-function $\varphi$, say
\begin{equation}
    \Psi(\vec{r}_{1},\vec{r}_{2},\vec{r}_{3})
    \Psi(\vec{r}_{4},\vec{r}_{5},\vec{r}_{6})
    \varphi(\vec{r}).
 \end{equation}
Perhaps the potential $V$ governing $\varphi(\vec{r})$ mimics the 
delicate dynamics to be expressed in a multichannel framework.  One 
might then be afraid that in the $\NNbar$ case, the distortion of the 
incoming bags $\Psi$ could be more pronounced.  In this case, the 
$G$-parity rule should be applied for each channel and for each 
transition potential separately, not a the level of the effective 
one-channel potential $V$.

{\em 4)} It would be very desirable to probe our theoretical ideas on 
the long- and intermediate-distance $\NNbar$ potential by detailed 
scattering experiments, with refined spin measurements to filter out 
the short-range contributions.  Unfortunately, only a fraction of the 
possible scattering experiments have been carried out at 
LEAR \cite{Martin:1993ij}, and uncertainties remain.  The available 
results are however compatible with meson-exchange models supplemented 
by annihilation.  The same conclusion holds for the detailed 
spectroscopy of the antiproton--proton atom \cite{Gotta:1999}.

\subsection{$\mathbf{\NNbar}$ spectra}
\label{spectra}
The first spectral calculations based on explicit $\NNbar$ 
potentials were rather crude.  Annihilation was first omitted to get a 
starting point, and then its effects were discussed qualitatively.  
This means the real part of the potential was taken as given by the 
$G$-parity rule, and regularised at short distances, by an empirical 
cut-off.  Once this procedure is accepted, the calculation is rather 
straightforward.  One should simply care to properly handle the 
copious mixing of $L=J-1$ and $L=J+1$ components in natural parity 
states, due to tensor forces, especially for isospin $I=0$ \cite{Dover:1978br}.

The resulting spectra have been discussed at length in Refs. \cite{Shapiro:1978wi,Buck:1979rt}. Of course, the number of bound 
states, and their binding energy increase when the cut-off leaves more 
attraction in the core, so no detailed quantitative prediction was 
possible. Nevertheless, a few qualitative properties remain when the 
cut-off varies: the spectrum is 
rich, in particular in the sector with isospin $I=0$ and natural 
parity corresponding to the partial waves  $\SLJ3P0$, 
$\SLJ3S1-\SLJ3D1$, $\SLJ3P2-\SLJ3F2$, corresponding to 
$J^{PC}I^{G}=0^{++}0^{+}$, $1^{--}0^{-}$, $2^{++}0^{+}$, respectively.
The abundant candidates for ``baryonium'' in the data available at 
this time \cite{Montanet:1980te} made this quasi-nuclear baryonium
 approach plausible \cite{Shapiro:1978wi,Dover:1979zj}.
 
 As already mentioned, annihilation was first neglected. Shapiro and his 
 collaborators \cite{Shapiro:1978wi} insisted on the short-range 
 character of annihilation and therefore claimed that it should not distort much the 
 spectrum. Other authors acknowledged that annihilation should be rather 
 strong, to account for the observed cross-sections, but should  
  affect mostly the S-wave states, whose binding rely on the 
 short-range part of the potential, and not too much the $I=0$, 
 natural parity states, which experience long-range tensor forces.
 
This was  perhaps a too optimistic view point. For instance, an 
explicit calculation \cite{Myhrer:1976by} using a complex optical 
potential fitting the observed cross-section showed that no $\NNbar$ 
bound state or resonance survives annihilation. In 
Ref. \cite{Myhrer:1976by}, Myhrer and Thomas used a brute-force 
annihilation. It was then argued that maybe annihilation is weaker, 
or at least has more moderate effects on the spectrum, if one accounts 
for  

{\em 1)} its energy dependence: it might be weaker below 
threshold, since the phase-space for pairs of meson resonances is more 
restricted.  It was even argued \cite{Green:1980pu} that part of the 
observed annihilation (the most peripheral part) in scattering 
experiments comes from transitions from $\NNbar$ scattering states to a 
$\pi$meson plus a $\NNbar$ baryonium, which in turn decays.  Of 
course, this mechanism does not apply to the lowest baryonium.

{\em 2)} its channel dependence: annihilation is perhaps less strong in 
a few partial waves. This however should be checked by fitting 
scattering and annihilation data.

{\em 3)} its intricate nature. Probably a crude optical model approach 
is sufficient to account for the strong suppression of the incoming 
antinucleon wave function in scattering experiments, but too crude for 
describing baryonium. Coupled-channel models have thus been developed 
(see, e.g., Ref. \cite{Carbonell:1991rw} and references therein). It 
turns out that in coupled-channel calculations, it is less difficult to
accommodate simultaneously large annihilation cross sections and 
relatively narrow baryonia.

\section{Multiquark baryonium}
\label{se:multi}
At the time where several candidates for baryonium were proposed, the 
quasi-nuclear approach, inspired by the deuteron described as a $\NN$ bound 
state, was seriously challenged by a direct quark picture.

Among the first contributions, there is the interesting remark by 
Jaffe  \cite{Jaffe:1977ig} that ${\q}^{2}{\qbar}^{2}$ S-wave are not 
that high in the spectrum, and might even challenge P-wave $\qqbar$ to 
describe scalar or tensor mesons.  From the discussions at this 
Conference, it is clear that the debate is still open.

It was then pointed out \cite{Jaffe:1978cv} that orbital excitations 
of these states, of the type $(\q^{2})$---$(\qbar^{2})$ have 
preferential coupling to $\NNbar$.  Indeed, simple rearrangement into 
two $\qqbar$ is suppressed by the orbital barrier, while the string can 
break into an additional $\qqbar$ pair, leading to $\q^{3}$ and 
$\qbar^{3}$.

Chan and collaborators \cite{Chan:1978qe,Chan:1978vw} went a little 
further and speculated about possible internal excitations of the 
colour degree of freedom. When the diquark is in a colour $\bar{3}$ 
state, they obtained a so-called ``true'' baryonium, basically 
similar to the orbital resonances of Jaffe. However, if the diquark 
carries a colour 6 state (and the antidiquark a colour 
$\bar{6}$), then the ``mock-baryonium'', which still hardly decays into 
mesons, is also reluctant to decay into $\N$ ad $\Nbar$, and thus is 
likely to be very narrow (a few MeV, perhaps). 

This ``colour chemistry'' was rather fascinating. A problem, however, 
is that the clustering into diquarks is postulated instead of being 
established by a dynamical calculation. (An analogous situation existed 
for orbital excitations of baryons: the equality of Regge slopes for 
meson and baryon trajectories is natural once one accepts that 
excited baryons consist of a quark and a diquark, the latter behaving 
as a colour $\bar{3}$ antiquark. The dynamical clustering of two of 
the three quarks in  excited baryons was shown only in 1985 \cite{Martin:1986hw}.)

There has been a lot of activity on exotic hadrons meanwhile, though 
the fashion focused more on glueballs and hybrids.  The pioneering bag 
model estimate of Jaffe and the cluster model of Chan et al.\ has been 
revisited within several frameworks and extended to new configurations 
such as ``dibaryons'' (six quarks), or pentaquarks (one antiquark, 
four quarks).  The flavour degree of freedom plays a crucial role in 
building configurations with maximal attraction and possibly more 
binding than in the competing threshold.  For instance, Jaffe pointed 
out that  (uuddss) might be more stable that two separated 
(uds) 
 \cite{Jaffe:1977yi}, and that such a stable dibaryon is more likely 
 in this strangeness $S=-2$ sector than in the  $S=-1$ or 
$S=0$ ones.  In the four-quark sector, configurations like 
$(\Q\Q\qbar\qbar)$ with a large mass ratio $m(\Q)/m(\qbar)$ are 
expected to resist spontaneous dissociation into two separated 
$(\Q\qbar)$ mesons (see, e.g., \cite{Richard:1994} and references 
therein).  For the Pentaquark, the favoured configurations 
$(\Q\qbar^{4})$ consist of a very heavy quark associated with light or 
strange antiquarks \cite{Gignoux:1987,Lipkin:1987sk}.

\section{Multiquark states vs.\ $\mathbf{\NNbar}$ states}
\label{se:multi-vs-QNB}
An obvious question is whether the picture of two hadrons interacting 
by exchanging mesons is more or less realistic than a direct 
approach using quark dynamics. One cannot give a general answer, as it 
depends on the type of binding one eventually gets for the state.

In the limit of strong binding, a multiquark system can be viewed as a 
single bag where quarks and antiquarks interact directly by exchanging 
gluons.  

For a multiquark close to its dissociation threshold, we have more 
often two hadrons experiencing their long-range interaction.  Such a 
state is called an ``hadronic molecule''.  There has been many 
discussions on such 
molecules \cite{Weinstein:1982gc,Close:1987aw,Barnes:1985cy,Dooley:1992bg,Ericson:1993wy,Manohar:1993nd,Tornqvist:1991ks,Tornqvist:1994ng,Barnes:1999hs}, 
$\hbox{K}\overline{\hbox{K}}$, $\hbox{D}\overline{\hbox{D}}$ or 
$\hbox{B}\hbox{B}^{*}$.  In particular, pion-exchange, due to its long 
range, plays a crucial role in achieving the binding of some 
configurations.  From this respect, it is clear that the baryonium 
idea has been very inspiring.

\section*{Acknowledgments}
\label{Ackn}
I would like to thank S.~Narison for the very stimulating 
atmosphere of this Conference, and S.U.~Chung and L.~Montanet for 
enjoyable discussions on exotic hadrons. 


\end{document}